\documentclass[pra,onecolumn,superscriptaddress]{revtex4}

\usepackage{float}
\usepackage{graphicx}
\usepackage{epsfig}
\usepackage{dcolumn}
\usepackage{amsmath,amssymb,amsthm}
\usepackage{bm}
\usepackage{verbatim}
\usepackage{natbib}
	
\usepackage{color}

\newcommand{\ket}[1]{|#1\rangle}

\newcommand{\ketbra}[2]{| #1 \rangle \langle #2 |}
\newcommand{\M}[1]{\mathcal{#1}}

\newcommand{\norminf}[1]{\parallel #1 \parallel_\infty}

\begin{document}
\title{Witnessed entanglement and the geometric measure of
 quantum discord} 
\author{Tiago Debarba} 
\email{debarba@fisica.ufmg.br}
\author{Thiago O. Maciel}
\author{Reinaldo O. Vianna}
\email{reinaldo@fisica.ufmg.br}

\affiliation{Departamento de F\'{\i}sica - ICEx - Universidade Federal de Minas Gerais,
Av. Pres.  Ant\^onio Carlos 6627 - Belo Horizonte - MG - Brazil - 31270-901.}

\date{\today}
\begin{abstract}
We establish relations between  geometric quantum
discord and  entanglement quantifiers obtained by means of optimal witness
operators. In particular, we prove a relation between negativity and geometric
discord in the Hilbert-Schmidt norm, which is slightly different from
a previous conjectured one
[1].
 We also show that, redefining the 
geometric discord with the trace norm, better bounds can be obtained.
We illustrate our results numerically.

\end{abstract}
\maketitle

\section{Introduction}
Entanglement has been widely investigated in the last years \cite{RMP-Horodecki},
 and is a resource that allows for tasks that cannot be performed classically,
as teleportation \cite{b93}, quantum key distribution \cite{ekert}, 
superdense coding \cite{bw92}, and  speed-up 
of some algorithms \cite{speedup},  just to cite a few examples. Therefore 
entanglement is an indisputable signature of the non-classicallity of a state.
Nevertheless, some authors have argued  that there is more to the
{\em quantumness} of a state than just its entanglement 
\cite{hv2001,oz2001, discordia-review}.
This notion of quantumness beyond entanglement is captured by the 
{\em quantum discord}, which is defined as all the correlations 
contained in a state but the classical ones \cite{hv2001}, or as a  measure
of disturbance of a state after local measurements \cite{oz2001}, both
definitions being compatible with a class of separable states with
non-null quantum discord. 

Recent investigations suggest that quantum discord can be considered
a resource that gives a quantum advantage \cite{discordia-recurso}. 
In order to deepen our understanding of both the usefulness of such a resource
and how {\em quantum} it really is, it is important to devise operational
means to quantify it, and also to relate it to quantum entanglement.
In this respect, the {\em geometric discord} \cite{dvb2010}  defined as
the distance between the state of interest and a properly 
defined {\em classical} state is an invaluable tool.
Unhappily the {\em classical states} do not form a convex set \cite{facca2010},
 and therefore
one cannot use the well known separating hyperplane theorem to characterize
discord  as is done with the witness operators in the case of the entanglement
problem \cite{RMP-Horodecki}.

Interesting investigations relating entanglement and discord have been
done recently \cite{ga2011, f2011}.  
In \cite{f2011}, entanglement of formation is related to discord in a 
conservation equation, and
in \cite{ga2011} geometric discord is conjectured to be bounded by the
negativity. 
While entanglement of formation is not computable in general, 
many other interesting entanglement quantifiers can be expressed in terms of 
optimal entanglement witnesses \cite{b2005} which, by its turn,
can be calculated numerically by means of efficient semidefinite programs
\cite{b2004,b2006}.  In this work we will explore bounds for geometric
discord by means of optimal entanglement witnesses. In particular, 
we will prove that negativity bounds the geometric discord.

This paper is organized as follows. In Sec.II we briefly revise quantum 
discord, and   propose a redefinition of geometric  discord using the Schatten 
$p$-norm. In Sec.III,  we recall the {\em witnessed entanglement}, with special
attention to both negativity and robustness.
In Sec.IV, we derive bounds for geometric discord using witnessed entanglement.
In Sec.V, we  illustrate our results for Werner states and some families
of bound entangled states. 
We conclude in Sec.VI.

\section{Quantum Discord}
\label{disc}

The total 
amount of correlations of a  bipartite system $AB$
is quantified by the well known mutual information, which in the classical
case can be written in two equivalent forms linked by Bayes' rule, namely:
$I(A:B) = H(A) + H(B) - H(AB) = H(A) - H(A|B)$, being $H(X)$ the Shanon entropy 
of $X$ and $H(X|Y)$ the conditional entropy of $X$ given $Y$.
For a quantum system $\rho_{AB}$, the mutual information is defined in
terms of the von Neuman entropy $S(\rho)=-Tr(\rho \log\rho)$, and reads:
\begin{equation}
\label{IAB}
I(\rho_{AB}) = S(\rho_A) + S(\rho_B) - S(\rho_{AB}),
\end{equation}
with $\rho_A$ and $\rho_B$  being the marginals of $\rho_{AB}$. However, 
the definition of a quantum {\em conditional entropy} is dependent
on the choice of a given POVM $\{\Pi_k\}$ to be measured on party B,
and the two expressions for the mutual information are no longer 
equivalent.
While Eq.\ref{IAB} still quantifies the total amount of
 correlations in the quantum 
state, the other expression involving the conditional entropy needs
 some attention.
After the measurement of $\Pi_k$ on $B$, party $A$ is left in the 
state $\rho_{A|k}=Tr_B(\mathbb{I}\otimes\Pi_k \rho_{AB})/p_k$, with probability
$p_k=Tr(\mathbb{I}\otimes\Pi_k \rho_{AB})$. Now we can write the 
conditional entropy associated to the POVM $\{\Pi_k\}$ as:
\begin{equation}
S(\rho_{A|B})=\sum_k p_k S(\rho_{A|k}).
\end{equation}
 $J(\rho_{AB},\{\Pi_k\})\equiv S(\rho_A)-S(\rho_{A|B})$ quantifies
the classical correlations contained in $\rho_{AB}$ under measurements
in the given POVM. Therefore, maximizing  $J(\rho_{AB},\{\Pi_k\})$ over
all POVMs quantifies the classical correlations in the
quantum state, namely \cite{hv2001,discordia-review} :
\begin{equation}
\label{JAB}
J_{AB}(\rho)=S(\rho_{A})-\min_{\Pi_{k}}\sum_{k}p_{k}S(\rho_{A|k}),
\end{equation}
where the POVMs can be chosen to be rank-one \cite{hkz2004}.
Finally,  the {\em quantum discord} is the disagreement between the
nonequivalent expressions of mutual information in the quantum case, 
namely \cite{oz2001, discordia-review}:
\begin{equation}
\label{DAB}
D(\rho_{AB})=I(\rho_{AB})-J(\rho_{AB}).
\end{equation}
Note that $D(\rho_{AB})$ is non-negative and asymmetric with respect 
to $A\leftrightarrow B$.

As discord is supposed to measure the quantumness of a state, it is no
 wonder that the maximally entangled states Eq.\ref{maxent} are the
 most discordant, while states which are a mere encoding of classical
probability distributions Eq.\ref{cc} are concordant (i.e. 
$I(\rho_{AB})=J(\rho_{AB})$) \cite{o3h2002}.
Bipartite maximally entangled states in 
$\mathcal{B}(\mathbb{C}^{d=d_A\times d_B})$, with $d_A=d_B$, have the form:
\begin{equation}
\label{maxent}
\phi=\frac{1}{d_A} \sum_{i,j=1}^{d_A}\ketbra{ii}{jj},
\end{equation}
while {\em classical states} can be written as:
\begin{equation}
\label{cc}
\xi=\sum_{i,j=1}^{d_A} p_{ij}\ketbra{e_i}{e_i}\otimes\ketbra{f_j}{f_j},
\end{equation}
where $\ket{e_i}$ and $\ket{f_j}$ are two  orthonormal bases. 
Note however, that as discord is asymmetric, if the measurements 
are to be done in subsystem $B$, the following class of states are also
concordant or classical:
\begin{equation}
\label{qc}
 \xi=\sum_{j=1}^{d_A} p_{j}\rho_{j}\otimes\ketbra{f_{j}}{f_{j}}.
 \end{equation}
To distinguish these two classes of classical states, sometimes
the former is referred to as {\em classical-classical}, while
the later is {\em quantum-classical}.

An alternative definition for quantum discord is based on the  distance 
between the given quantum state and the closest classical state
 \cite{dvb2010, mpsvw2010, bggfcz2011, lf2010,bm2011}.
Adopting the  Hilbert-Schmidt norm $\Vert \Vert_{(2)}$,
 we can write \cite{dvb2010}:
\begin{equation}
\label{D2}
D_{(2)}(\rho_{AB})=\min_{\xi\in\Omega}\Vert\rho-\xi\Vert^{2}_{(2)},
\end{equation}
where $\Omega$ is the set of  zero-discord states. This measure can 
be interpreted as the minimal disturbance after local measurements on 
subsystem $B$. In  this case $\Omega$ contains the states in Eq.\ref{qc},
and $D_{(2)}$ can be calculated analytically for some states 
\cite{lf2010}.

Consider the  Schatten $p$-norm for some matrix $A$  and positive 
integer $p$:
\begin{equation}
\Vert A\Vert_{(p)}=\{Tr[(A^{\dag}A)^{p/2}]\}^{1/p},
\end{equation}
which, for $p=2$ is the Hilbert-Schmidt norm.
In finite Hilbert spaces, these norms induce the same ordering \cite{zyc}.
Therefore we propose to extend the geometric discord for any
$p$-norm, namely:
\begin{equation}\label{1}
 D_{(p)}(\rho_{AB})=\min_{\xi\in\Omega}\Vert\rho_{AB} - \xi \Vert^{p}_{(p)}.
\end{equation}
Note that for $p\ge q$, we have $\Vert A\Vert_{p}\leq\Vert A\Vert_{q}$.
It follows that  the 1-norm is the most distinguishable distance in 
Hilbert space.
Therefore,  we shall investigate the geometric discord in the 
1-norm ($D_{(1)}$), besides the usual $D_{(2)}$. As we shall see,
it is easy to bound  these geometric discords by entanglement
witnesses.

\section{Witnessed Entanglement}
\label{witnessed.entanglement}

Entanglement witnesses  are Hermitian operators (observables - $W$) whose 
expectation values contain information about the entanglement of 
quantum states. 
The operator $W$ is an entanglement witness for a given entangled 
quantum state $\rho$ if the following conditions are satisfied 
\cite{hhh1996}: its 
expectation  value is negative for the particular entangled quantum state 
($Tr(W\rho) < 0 $), while it is non-negative  on the set of separable 
states ($S$) 
($\forall \sigma \in \M{S}, \,\,\, Tr(W\sigma)\geq 0 $).
We are particularly  interested in  optimal entanglement witnesses. 
\textit{$W_{opt}$} is the OEW
for the state $\rho$ if 
\begin{equation}
Tr(W_{opt}\rho) = \min\limits_{W \in \M{M}} \,Tr(W\rho),
\label{optimal.witness.definition}
\end{equation}
where $\M{M}$ represents a compact subset of the set of 
entanglement witnesses $\M{W}$ \cite{b2005}.

OEWs can be used to quantify  entanglement.
Such quantification is related to  the choice of the set $\M{M}$, where 
different sets will determine different quantifiers \cite{b2005}. 
We can define these quantifiers by:
\begin{equation}
E_w(\rho) = \max{(0 , -\min\limits_{W \in \M{M}} \,Tr(W\rho))}.
\label{OEW}
\end{equation}

An example of a quantifier that can be calculated using OEWs is the 
 Generalized Robustness of entanglement \cite{vt1999} ( $\M{R}_g(\rho)$), 
which is defined as the minimum required mixture such that 
a separable state is obtained.
Precisely, it is the minimum value of $s$ such that
\begin{equation}
\label{gr}
\sigma=\frac{\rho + s\varphi}{1+s}
\end{equation} 
be a separable state, where $\varphi$ can be any state. 
We know that the Generalized Robustness can be calculated from 
Eq.\ref{OEW}, using $\M{M} = \{W \in \M{W}\, |\, W \leq \mathbb{I}\}$
\cite{b2005}, 
where  $\mathbb{I}$ is the identity operator; 
in other words,
\begin{equation}
\M{R}_{g}(\rho) = \max{(0, -\min_{\{W \in \M{W}\, |\, W \leq \mathbb{I}\}} Tr(W\rho))}.
\end{equation} 
A particular case of the Generalized Robustness is the Random Robustness,
where $\varphi$ in Eq.\ref{gr} is taken to be the maximally mixed state
($\mathbb{I}/d$). In this case, the compact set of entanglement witnesses is 
$\mathcal{M}=\{ W\in\mathcal{W}|Tr(W)=1 \}$. The Random Robustness 
$\M{R}_r(\rho)$ quantifies the resilience of the entanglement to 
white noise, and is given by \cite{b2006}:
\begin{equation}
\label{rr}
d\times\M{R}_{r}(\rho) = \max{(0, -\min_{\{W \in \M{W}\, |\, Tr(W)=1 \}} Tr(W\rho))}.
\end{equation}

The well known Negativity for bipartite states, which
is the sum of the negative eigenvalues of the partial transpose of
the given state, $ \M{N}(\rho)\equiv(\Vert\rho^{T_A}\Vert_{(1)}-1)/2$,
  can also be expressed in terms of OEWs as \cite{b2005}:
\begin{equation}
\label{WN}
\mathcal{N}(\rho)=\max\{ 0, - \min_{0\leq W^{T_{A}} \leq 
\mathbb{I}} Tr(W\rho)  \}.
\end{equation}

The construction of entanglement witnesses is a hard problem.
In an interesting  method proposed 
by Brand\~ao and Vianna   \cite{b2004},  the optimization of 
entanglement  witnesses  is cast as a
{\em robust semidefinite program} (RSDP). 
Despite  RSDP is  computationally intractable, it is possible to 
perform a probabilistic relaxation turning it into   a semidefinite
program(SDP), which can be solved efficiently \cite{convexoptimization}.

\section{Bounding geometric discord with witnessed entanglement}
\label{theorem}

In this section we show that geometric discord, in any norm, is lower
bounded by entanglement. In particular,  we show that norm-2 geometric discord
is bounded by negativity, but the relation is slightly different from
that conjectured by Girolami and Adesso [12].

For any  two operators $A$,$B$$\in\mathcal{B}(\mathbb{C}^{d}) $ 
and the Schatten $p$-norm $\Vert A \Vert_{p} = Tr[(AA^{\dag})^{p/2}]^{1/p}$, 
the following inequality holds:
\begin{equation}\label{sec4.1}
\Vert A\Vert_{p}\Vert B\Vert_{q}\geq|Tr[AB^{\dag}]|,  
\end{equation} 
where $1/q+1/p = 1$. 

The geometrical discord for a state 
$\rho\in\mathcal{B}(\mathbb{C}^{d=d_A\times d_B})$ is:
\begin{equation} \label{sec4.2}
D_{p}(\rho) = \Vert \rho - \bar{\xi} \Vert_{p}^{p},  
\end{equation}
where $\bar{\xi}$ is the closest non-discordant state. 
The witnessed entanglement of $\rho$ can be written as:
\begin{equation} \label{sec4.3}
E_{w}(\rho) = \min\{0, - Tr[W_{\rho}\rho] \},
\end{equation} 
where  $W_{\rho}$ is the optimal entanglement witness of 
$\rho$. Plugging  
$A= \Vert \rho - \bar{\xi} \Vert_{p}^{p}$  and 
$B = W_{\rho}$ in Eq.\ref{sec4.1}, we get:
\begin{equation} \label{sec4.4}
\Vert \rho - \bar{\xi} \Vert_{p} \Vert W_{\rho} \Vert_{q} 
\geq |Tr[(\rho -\bar{ \xi})W_{\rho}]|.
\end{equation}
If $\rho$ is entangled and  $\bar{\xi}$ is separable 
we have \newline
$|Tr[(\rho -\bar{ \xi})W_{\rho}]|\geq|Tr[\rho W_{\rho}]|$, 
thus:
\begin{equation}\label{sec4.5}
\Vert \rho - \bar{\xi} \Vert_{p} 
\geq \frac{|Tr[\rho W_{\rho}]|}{ \Vert W_{\rho} \Vert_{q}},
\end{equation}
which in terms of geometric discord (Eq.2) reads:
\begin{equation}\label{sec4.6}
D_{(p)}(\rho) \geq \Bigg(\frac{E_{w}(\rho)}{\Vert W_{\rho}\Vert_{q}}\Bigg)^{p}.
\end{equation}
Therefore, given any entanglement witness (it does not need to be optimal),  we have a bound for
geometric discord in any norm. Note that Eq.6 is also valid for 
multipartite states. For norm-1 and norm-2, Eq.6 reduces to:
\begin{equation}\label{sec4.7}
 D_{(1)}(\rho) \geq \frac{E_{w}(\rho)}{\Vert W_{\rho}\Vert_{\infty}},
 \end{equation}

\begin{equation}\label{sec4.8}
 D_{(2)}(\rho) \geq \frac{E_{w}^{2}(\rho)}{Tr [W_{\rho}^{2}]}.
 \end{equation}

If $W_\rho$ is the entanglement witness for the negativity,
$\M{N}(\rho)=E_w(\rho)$ (see Eq.\ref{sec4.3}), then
 $Tr[W_{\rho}^2]  =n_{-}$, 
where $n_-$ is the number of negative eigenvalues of the partial
transpose of $\rho$ ($\rho^{T_A}$). 
Thus, in norm-2, discord is lower bounded by negativity as:
\begin{equation}\label{sec4.9}
D_{(2)}(\rho) \geq \frac{\mathcal{N}^{2}(\rho)}{n_{-}},
\end{equation}
where $0< n_{-}\leq d-1$ (remember $d=d_A\times d_B$).
For norm-1 discord, one has to calculate $\Vert W_{\rho}\Vert_{\infty}$,
which is simply the largest eigenvalue of $W_\rho$ in absolute value, 
 and use Eq.\ref{sec4.7}. Note that it is easy. One has just  to form a rank-$n_-$ 
projector with the eigenstates or $\rho^{T_A}$ associated to the 
$n_-$ negative eigenvalues, then $W_\rho$ is the partial transpose
of this projector.

\section{Numerical applications}
\label{numerical}
In this section we illustrate our results with numerical calculations on 
maximally entangled pure staes, Werner states and bound entangled states. 
We consider the negativity and random robustness.

\begin{figure}
\centering 
\epsfig{figure=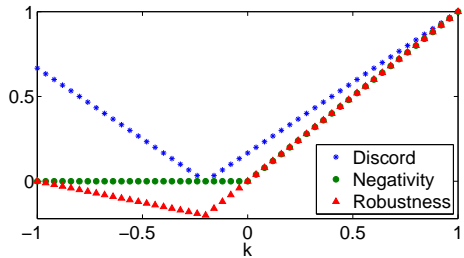, width=8.0cm} 

\caption{(Color online) 1-norm geometric discord, negativity and 
random robustness for 
$5\otimes 5$ 
Werner states.}
\label{f1}
\end{figure}

\subsection{Werner states}
Werner states  ($d_A\otimes d_A$) \cite{werner}, which are of the Bell diagonal type, 
can be written as:
\begin{equation}
\rho_{w}=\frac{d_A +k}{d_A^3-d}\mathbb{I}_{d}-\frac{d_A k-1}{d_A^3-d_A}\ketbra{\psi^{+}}{\psi^{+}},
\end{equation} 
where $\ket{\psi^{+}}= \sum_{i,j=0}^{d_A -1}(\ket{ij}+\ket{ji})$. The parameter
$k$ is in the interval $[-1,1]$, and the state is entangled for $k>0$.

In (Fig.\ref{f1}) we compare 
negativity, random robustness and 1-norm geometric discord. 
Note that negativity and random robustness coincide in the
entangled region and are always less than the discord. 
Note also that the random robustness in the non-entangled region 
($Tr(\rho_w W_{\rho_w})\geq 0$) has a functional behavior similar
to the discord.

\subsection{Bound-entangled states}
Bound entangled states have positive partial transpose ($ppt$) and are
known to be undistillable \cite{3h98}. The negativity is useless in 
this case, but the random robustness can give an interesting bound for
the discord.

\subsubsection{Horodecki's $ppt$-entangled states}
Consider a Hilbert space $\mathbb{C}^{3}\otimes\mathbb{C}^{3}$, and 
a canonical orthonormal basis  $\{\ket{i}\}_{0,1,2}$. 
Take the following three states \cite{h97,3h98}:
\begin{equation}
Q=\mathbb{I}\otimes\mathbb{I}-\big[ \sum_{i=0}^2 \ketbra{i}{i}\otimes \ketbra{i}{i} + \ketbra{2}{2}\otimes\ketbra{0}{0} \big],
\end{equation}
\begin{equation}
\ket{\psi}=\frac{1}{3} \big[ \ket{0}\ket{0}+\ket{1}\ket{1}+\ket{2}\ket{2} \big]
\end{equation}
and
\begin{equation}
\ket{\phi_{k}}=\ket{2}\otimes\Big[ \sqrt{\frac{1+k}{2}}\ket{0}+ \sqrt{\frac{1-k}{2}}\ket{2}  \Big],
\end{equation}
where $0\leq k \leq 1$. 
The following convex combination is a 
$ppt$-entangled state for $0 \leq k < 1$, and is separable for $k=1$:
\begin{equation}\label{h}
\varrho_{k}= \frac{k}{8k+1} \big[ 3 \ketbra{\psi}{\psi} +Q  \big] + \frac{1}{8k+1} \ketbra{\phi_{k}}{\phi_{k}}.
\end{equation}
Note, in Fig.2a, that  the most discordant state of this family is
 the less
entangled one, and vice-versa. However, it is not a general characteristic
of bound entangled states, as can be seen in the next example (Fig.2b).

\subsubsection{UPB entangled  states }
In a bipartite Hilbert space $\mathcal{H}=\mathbb{C}^{d_A}\otimes\mathbb{C}^{d_B}$, 
an orthogonal product basis (PB)
is an $l$-dimensional set of separable states  spanning a subspace $\mathcal{H}_l$ of $\mathcal{H}$. 
When the complement of this PB in  $\mathcal{H}$ has only entangled states,
we say that the complete basis containing PB is 
a unextendible product basis (UPB)
 \cite{upbprl}.

Consider the following  three classes of vectors in  $\mathcal{H}=\mathbb{C}^{4}\otimes\mathbb{C}^{4}$:
\begin{eqnarray*}
\ket{v_{j}}&=&\ket{j}\otimes\frac{\ket{(j+1)\mod4}-\ket{(j+2)\mod4}}{\sqrt{2}}, \\
\ket{u_{j}}&=&\frac{\ket{(j+1)\mod4}-\ket{(j+2)\mod4}}{\sqrt{2}}\otimes\ket{j}, \\
\ket{w} &=& \frac{1}{4}\sum_{i,j=0}^3\ket{i}\otimes\ket{j}. 
\end{eqnarray*}
Now define the following vectors:
$\ket{\psi_{k}}=\ket{v_{k}}$ for $k=0,1,2,3$, 
$\ket{\psi_{k}}=\ket{u_{k\mod4}}$ for $k=4,5,6,7$,
 and $\ket{\psi_{k}}=\ket{w}$ for $k=8$. 
Finally, a $ppt$-entangled state is given by:
\begin{equation}\label{UPB}
\rho=\frac{1}{7} \Big(\mathbb{I} - \sum_{k=0}^8\ketbra{\psi_{k}}{\psi_k} \Big).
\end{equation}
In (Fig.2b), we plotted random robustness and 1-norm geometric discord for
the  convex mixture:
\begin{equation}
\sigma =    \frac{s}{16}\mathbb{I}+ (1-s)\rho,
\end{equation}
which is separable  for $s> 0.169$.  
We see that the discord is always greater than the entanglement, and 
the most discordant states are also the most entangled.

\begin{figure}
\centering 
\epsfig{figure=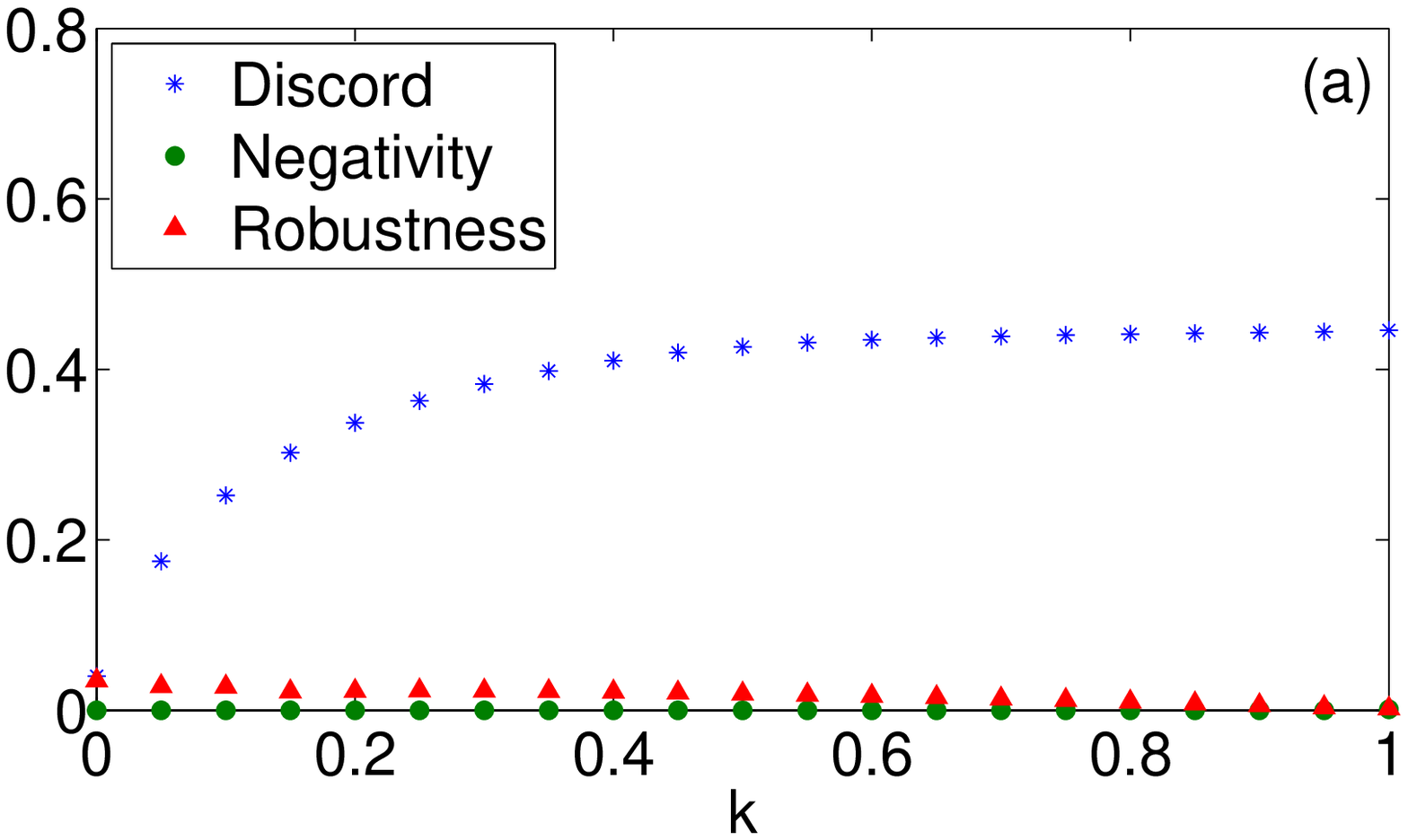, width=8.0cm}
\epsfig{figure=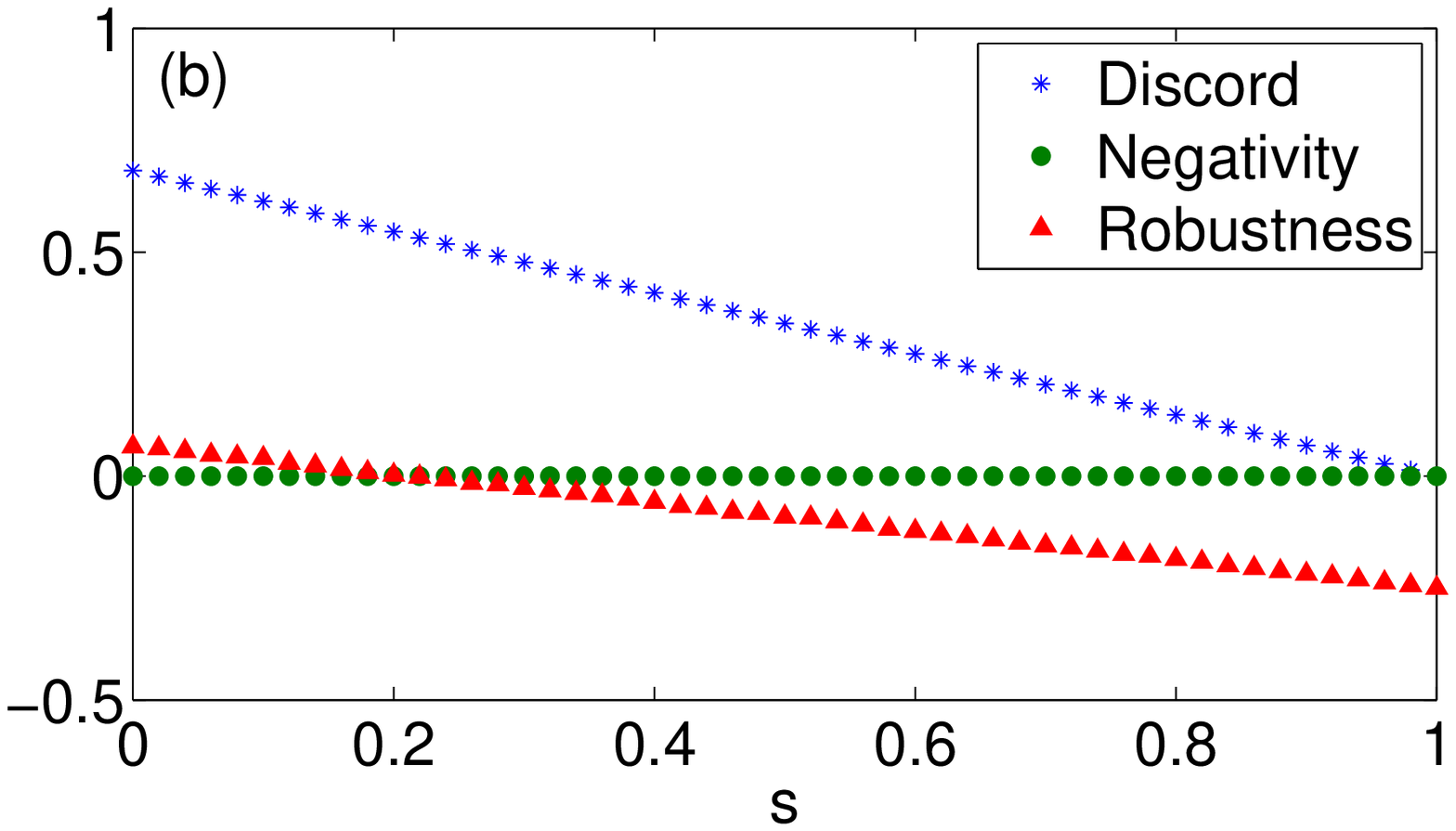, width=8.0cm} 

\caption{(Color online) 1-norm geometric discord and random robustness for 
$3\otimes 3$ (a) and $4\otimes 4$ (b) $ppt$-entangled states.}
\label{f2}
\end{figure}

\subsection{Pure states versus Werner states}

In Tabs. I and II, we compare the bounds for norm-1 and norm-2 in 
two maximally entangled states, an $8\otimes 8$ Werner state, and
 a $2\otimes 8$ mixed state whose density matrix coincides with
the $8\otimes 8$ Werner state. The norm-1 bounds are much better
than the norm-2 ones. Tab.III is an {\em Erratum} for the published
version of this paper.

\begin{table}
\begin{tabular}{||c|r|r|c|r|r|r||} \hline
	&  $Tr(\rho {W_n})$ & $Tr(\rho {W_r})$ & $ Tr({W_n}^2) $ & $ Tr({W_r}^2) $ &
        $\norminf{W_n}$ &  $\norminf{W_r} $
            \\ \hline
$(d=2\otimes 2)\,\rho=\ketbra{\Phi^+}{\Phi^+}  $ & -0.5000  & -0.5000   & 1  & 1.0000  & 0.5000  & 0.5000 \\    \hline
$(d=4\otimes 4)\,\rho=\ketbra{\Phi^+}{\Phi^+}  $ & -1.5000  & -0.2500   & 6  & 0.1677  & 1.5000   &  0.2503 \\    \hline
$(d=8\otimes 8)\,\rho= \rho_{w}(8,-1)  $         & -0.1250  & -0.1250   & 1  & 1.0000  & 0.1250  & 0.1250 \\    \hline
$(d=2\otimes 32)\,\rho=\rho_{w}(8,-1)  $         & -0.1786  & -0.0179   & 10  & 0.1013  & 0.5000  & 0.0600 \\    \hline
\end{tabular}
\caption{Entanglement and some properties of the corresponding entanglement witness.
$W_n$ and $W_r$ are the entanglement witnesses for negativity and random robusteness,
respectively.}
\end{table}

\begin{table}
\begin{tabular}{||c|r|r|r|r|r|r||} \hline
	& $D_2$ & $\frac{Tr(\rho {W_n})^2}{Tr({W_n}^2)}$ & $\frac{Tr(\rho {W_r})^2}{Tr({W_r}^2)}$ &
          $D_1$ & $\frac{-Tr(\rho {W_n})}{\norminf{W_n}}$ &  $\frac{-Tr(\rho {W_r})}{\norminf{W_r}}$ 
            \\ \hline
$(d=2\otimes 2)\,\rho=\ketbra{\Phi^+}{\Phi^+}  $ & $0.5000$  & $0.2500$   &  0.2500 & 1.0000    & 1.0000   & 1.0000 \\    \hline
$(d=4\otimes 4)\,\rho=\ketbra{\Phi^+}{\Phi^+}  $ & $0.7500$  & $0.3750$   & 0.3727  & $1.5000$    & 1.0000   & 0.9988 \\    \hline
$(d=8\otimes 8)\,\rho= \rho_{w}(8,-1)  $         & $0.0179$  & $0.0156$   &  0.0156 & $1.0000$  &  1.0000  & 1.0000 \\    \hline
$(d=2\otimes 32)\,\rho=\rho_{w}(8,-1)  $         & $0.0102$  & $0.0032$   & $0.0032$  & $0.5714$   &$0.3580$  & $ 0.2983$ \\    \hline
\end{tabular}
\caption{Bounding geometric discord with witnessed entanglement.}
\end{table}

\begin{table}
\begin{tabular}{||c|r|r|r|r|r||} \hline
	& $D_2$ & $Eq.21=\frac{\M{N}^2}{d-1}$ & 
          $D_1$ & $Eq.27=\frac{\M{N}}{d}$ & $Eq.28=\frac{\M{R}_r}{d}$ 
           \\  \hline
$  (d=2\otimes 2)\,\rho=\ketbra{\Phi^+}{\Phi^+}          $ & $0.5000$   & $0.0833$   & $1.0000$  & $0.1250$  &  0.5000    \\  \hline
$  (d=4\otimes 4)\,\rho=\ketbra{\Phi^+}{\Phi^+}          $ & $0.7500$   & $0.1500$   & $1.5000$  & $0.0938$  &  0.2500     \\  \hline
$  (d=8\otimes 8)\,\rho= \rho_{w}(8,-1)          $         & $0.0179$   & $0.0002$   & $1.0000$  & $0.0020$  & 0.0020     \\  \hline
$   (d=2\otimes 32)\,\rho=\rho_{w}(8,-1)         $         & $0.0102$   & $0.0005$   & $0.5714$  & $0.0028$  & $0.0179$     \\  \hline
\end{tabular}
\caption{{\em Errata} for Eqs. 21, 27 and 28 in [Phys. Rev. A {\bf 86}, 024302 (2012)]}
\end{table}

\section{Conclusion}

We obtained bounds for geometric discord, in any norm,  in terms of  entanglement
witnesses (EW). Many known measures of entanglement can be expressed
by Optimal EWs, which implies that our bounds are quite general.
We note that, in a previous work \cite{fernando2012}, we showed how to 
calculate entanglement and geometric discord in systems
of indistinguishable fermions, and we checked that
the geometric discord was also bounded by the witnessed entanglement
in that case.

{\em Acknowledgments} - We thank Fernando G.S.L. Brand\~ao
for the discussions. Financial support by the
Brazilian agencies  FAPEMIG, CNPq, and  INCT-IQ (National
Institute of Science and Technology for Quantum Information).

\end{document}